\documentstyle[floats,multicol,aps,epsf,epsfig,prb]{revtex}
\newcommand\dg{^{\dag}}

\newcommand \etal {{\sl et al.}}

\def\3he{{$^3${\rm He}}}

\def\etal{{\it et al.}}

\def\slD{\raise.15ex\hbox{$/$}\kern-.57em\hbox{$D$}}
\def\dsl{\raise.15ex\hbox{$/$}\kern-.57em\hbox{$\Delta$}}
\def\slp{{\raise.15ex\hbox{$/$}\kern-.57em\hbox{$\partial$}}}
\def\nsl{\raise.15ex\hbox{$/$}\kern-.57em\hbox{$\nabla$}}
\def\sla{\raise.15ex\hbox{$/$}\kern-.57em\hbox{$\rightarrow$}}
\def\slla{\raise.15ex\hbox{$/$}\kern-.57em\hbox{$\lambda$}}
\def\gtwid{\raise.3ex\hbox{$>$\kern-.75em\lower1ex\hbox{$\sim$}}}
\def\ltwid{\raise.3ex\hbox{$<$\kern-.75em\lower1ex\hbox{$\sim$}}}

\def\12{{1\over2}}

\def\part{\partial}

\def\bethlogo{\vbox{\bf \line{\hrulefill} 
    \kern-.5\baselineskip 
    \line{\hrulefill\phantom{ ELIZABETH A. MASON }\hrulefill} 
    \kern-.5\baselineskip 
    \line{\hrulefill\hbox{ ELIZABETH A. MASON }\hrulefill} 
    \kern-.5\baselineskip 
    \line{\hrulefill\phantom{ 1411 Chino Street }\hrulefill} 
    \kern-.5\baselineskip 
    \line{\hrulefill\hbox{ 1411 Chino Street }\hrulefill} 
    \kern-.5\baselineskip 
    \line{\hrulefill\phantom{ Santa Barbara, CA 93101 }\hrulefill} 
    \kern-.5\baselineskip 
    \line{\hrulefill\hbox{ Santa Barbara, CA 93101 }\hrulefill}
    \kern-.5\baselineskip 
    \line{\hrulefill\phantom{ (805) 962-2739 }\hrulefill} 
    \kern-.5\baselineskip 
    \line{\hrulefill\hbox{ (805) 962-2739 }\hrulefill}}}
\def\lisalogo{\vbox{\bf \line{\hrulefill} 
    \kern-.5\baselineskip 
    \line{\hrulefill\phantom{ LISA R. GOODFRIEND }\hrulefill} 
    \kern-.5\baselineskip 
    \line{\hrulefill\hbox{ LISA R. GOODFRIEND }\hrulefill} 
    \kern-.5\baselineskip 
    \line{\hrulefill\phantom{ 6646 Pasado }\hrulefill} 
    \kern-.5\baselineskip 
    \line{\hrulefill\hbox{ 6646 Pasado }\hrulefill} 
    \kern-.5\baselineskip 
    \line{\hrulefill\phantom{ Santa Barbara, CA 93108 }\hrulefill} 
    \kern-.5\baselineskip 
    \line{\hrulefill\hbox{ Santa Barbara, CA 93108 }\hrulefill}
    \kern-.5\baselineskip 
    \line{\hrulefill\phantom{ (805) 962-2739 }\hrulefill} 
    \kern-.5\baselineskip 
    \line{\hrulefill\hbox{ (805) 962-2739 }\hrulefill}}}

\def\low#1{\lower.5ex\hbox{${}_#1$}}
\def\ltwid{\raise.3ex\hbox{$<$\kern-.75em\lower1ex\hbox{$\sim$}}}

\def\psl{\raise.15ex\hbox{$/$}\kern-.57em\hbox{$\partial$}}
\def\partt{\raise.15ex\hbox{$\widetilde$}{\kern-.37em\hbox{$\partial$}}}
\def\parts{\raise.15ex\hbox{$/$}{\kern-.6em\hbox{$\partial$}}}
\def\nablas{\raise.15ex\hbox{$/$}{\kern-.6em\hbox{$\nabla$}}}
\def\oprod{\hbox{$\rm O$}{\kern -0.8em\hbox{$\Pi$}}}
\def\partw#1{\raise.15ex\hbox{$/$}{\kern-.6em\hbox{${#1}$}}}

\def\gtappr{{{\lower4pt\hbox{$>$} } \atop \widetilde{ \ \ \ }}}
\def\ltappr{{{\lower4pt\hbox{$<$} } \atop \widetilde{ \ \ \ }}}

\def\topppageno1{\global\footline={\hfil}\global\headline
={\ifnum\pageno<\firstpageno{\hfil}\else{\hss\twelverm --\ \folio
\ --\hss}\fi}}

\def\toppageno2{\global\footline={\hfil}\global\headline
={\ifnum\pageno<\firstpageno{\hfil}\else{\rightline{\hfill\hfill
\twelverm \ \folio
\ \hss}}\fi}}

\def\relbd{\mathrel{{\bf\smash{{\phantom- \above1pt \phantom-
}}}}}

\def\ltdash{\raise-1.8pt\hbox{$\scriptscriptstyle |$}}

\def\dg{{^
{\dag}}}

\def\1{{\bf 1}}
\def\2{{\bf 2}}

\def\vk{\vec k}

\def\ell{{\it l } {\rm n}}

\def\cx2{\sqrt{c^2_x+c^2_y}}

\def\gkk{\gamma _{\vec k}}
\def\gk2{\gkk ^2}

\def\gtappr{{{\lower4pt\hbox{$>$} } \atop \widetilde{ \ \ \ }}}
\def\ltappr{{{\lower4pt\hbox{$<$} } \atop \widetilde{ \ \ \ }}}

\def\pbar{{\partial\kern-1.2ex\raise0.25ex\hbox{/}}}
\def\dg{{^{\dag}}}

\def\1{{\bf 1}}
\def\2{{\bf 2}}

\def\vk{\vec k}

\def\ell{{\it l } {\rm n}}

\def\cx2{\sqrt{c^2_x+c^2_y}}

\def\gkk{\gamma _{\vec k}}
\def\gk2{\gkk ^2}
\def\gtappr{{{\lower4pt\hbox{$>$} } \atop \widetilde{ \ \ \ }}}
\def\ltappr{{{\lower4pt\hbox{$<$} } \atop \widetilde{ \ \ \ }}}

\def\thickra{\hbox{\raise0.2pt\hbox{{$\bf >\mkern-13mu>\mkern-13mu>$}}}}
\def\thickrarrow{\hbox{\raise0.28pt\hbox{{$\bf >\mkern-13mu>\mkern-13mu>$}}}}

\begin{document}
%%%% the next line was added by xxx admin to ensure that
%%%% this paper doesn't use hypertex because of problems with references
\ifx\href\undefined\else\errmessage{hyperTeX disabled by xxx admin}\fi
\draft
% This is for two column
\twocolumn[\hsize\textwidth\columnwidth\hsize\csname
@twocolumnfalse\endcsname
\title{Theories of non-Fermi liquid behavior in heavy fermions.}
\author{ P. Coleman}
\address{ Serin Laboratory, Rutgers University,
P.O. Box 849, Piscataway, NJ 08855-0849, USA.}
\date{\today}
\maketitle
\begin{abstract}
I will review our incomplete understanding of non-Fermi liquid
behavior in heavy fermion systems at a quantum critical point.
General considerations suggest that critical
antiferromagnetic fluctuations do not destroy the Fermi surface by
 scattering the heavy electrons- but by actually breaking up the
internal structure of the heavy fermion.  I  contrast the weak,
and strong-coupling view of the quantum phase transition,
emphasizing puzzles and questions that recent experiments
raise.
\end{abstract}
\pacs{To be published in Proceedings of SCES '98, Paris, France.
}
\vskip2pc]

\def\beq{\begin{equation}}
\def\eeq{\end{equation}}
\def\bea{\begin{eqnarray}}
\def\eea{\end{eqnarray}}
\def \vk{\vec k}

Although heavy fermion behavior came to light over twenty years ago, we forget that consensus about the Fermi liquid
nature of the normal state took nearly a decade to develop. 
The last five years have seen a new struggle to understand the persistent
deviations from Fermi liquid behavior that occur in some of these compounds.  
\cite{sbarb}

Several aspects of non-Fermi liquid behavior will be discussed at this
conference, in particular, the vital effects of disorder and inhomogeniety.
\cite{bernal,miranda,castroneto} Here however,  I shall   focus on
a new body of  evidence which shows that non-Fermi liquid
behavior develops in single-crystal heavy fermion compounds at an antiferromagnetic
quantum critical point (Q.C.P).
Our understanding of 
the new metallic state this gives rise to  is currently  very incomplete and I
shall  try to emphasize the questions we are  posed.

Antiferromagnetic quantum critical behavior occurs in many heavy fermion
metals, including 
$CePd_2Si_2$,\cite{mathur} $CeIn_3$,\cite{mathur}
$CeNi_2Ge_2$,\cite{steglich} $CeCu_{6-x}R_x$ ($R= Au, 
\ Ag$),\cite{vonlohn,heuser} $CeRu_2Si_2$\cite{jaccard} and $U_2Pt_2In_2$.\cite{estrella}  
Either by pressure,\cite{julian} doping or a magnetic field,\cite{heuser} these compounds
can be reproducibly brought to a QCP
where the Ne\'el temperature vanishes.  
At this point  various non-Fermi liquid
properties develop, such as
\begin{itemize}

\item Anomalous power-law temperature dependence of the resistivity.
In $CePd_2Si_2$ for example, a powerlaw $\rho \propto T^{1.2}$ is seen on both the antiferromagnetic,
and the paramagnetic side of the transition.  

\item Non Curie temperature dependence of the susceptibility.

\item Anomalous logarithmic temperature dependence of the specific heat
$C_v/T =A{\rm ln}(T^*/T)$ . In
$CeCu_{6-x}R_x$, the same $A$ and $T^*$ are 
observed at the critical point, no matter how it is approached.\cite{vonlohn}

\end{itemize}
As the quality of the samples improves, the non-Fermi liquid behavior
persists, and in the case of 
$CePd_2Si_2$ and $CeGe_2Ni_2$,
superconductivity is found to develop at low temperatures in the
vicinity of the QCP.

\begin{figure}[here]
% ********   This is for two columns
\epsfxsize=3.0in 
% ***********For one column  ********************
%\epsfxsize=4.0in 
% ***********************************
\centering{\epsfig{figure=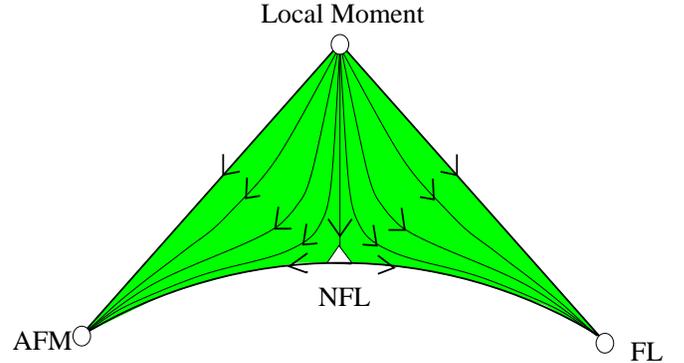,width=\linewidth}}
\vskip 0.1truein
\protect\caption{Schematic scaling diagram for the Kondo Lattice
}
\label{Fig1}
\end{figure}

Doniach\cite{doniach} first  proposed the possibility of dense Kondo
behavior in heavy fermion compounds,  suggesting that when the  ratio 
of the single ion Kondo temperature $T_K$ to the RKKY 
interaction scale $T_{RKKY}$ exceeds a 
critical value, magnetism would vanish. The possibility that the
transition might be continuous  was not appreciated until
recently, and has special
significance.\cite{tsvelik,continentino} If we  think of local moment 
antiferromagnetism (AFM)
and   Fermi liquid (FL)  behavior as two   competing attractive
fixed points of a renormalization group trajectory, then the existence 
of an antiferromagnetic QCP tells us that
these two limits are linked by new fixed point.
 (Fig. \ref{Fig1}).
\noindent 
As the temperature is lowered, the effective
Hamiltonian 
evolves away from high-temperature local moment behavior
to  one of  \underline{two} alternate attractive fixed points.
By tuning a material, using pressure or some other means, to the
critical value of 
$T_K/T_{RKKY}$, it is forced to evolve along a separatrix
to the QCP. 
More importantly, a
wide range of materials close to this critical value will pass close to the new
fixed point,  and over a large temperature range their properties, excitations
and interactions 
will be dominated by the  physics of this QCP.  
  
\subsection{ Weak versus strong-coupling approaches}

The scaling diagram tells us that there
are two ways of regarding the transition from
heavy fermion behavior to antiferromagnetism.  A ``weak coupling''
approach starts on the Fermi liquid side, and
regards the QCP as a
magnetic instability of the Fermi surface. The alternative strong coupling
approach starts from the magnetic side, and regards these metals as local moment
systems  which lose their magnetism once
the single-ion Kondo temperature is large enough to develop a dense Kondo
effect.  
\vskip 0.2truein
\begin{figure}[here]
% ********   This is for two columns
%\epsfxsize=3.0in 
% ***********For one column  ********************
%\epsfxsize=4.0in 
% ***********************************
\centering{\epsfig{figure=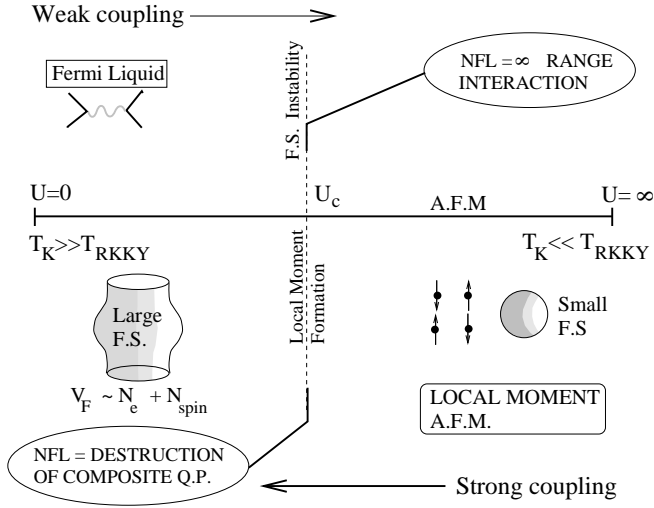,width=\linewidth}}
\vskip 0.1truein
\protect\caption{Contrasting the weak, and strong-coupling picture of
an antiferromagnetic QCP.
}
\label{Fig3}
\end{figure}
\noindent 

From the first perspective, the magnetic instability develops
in momentum space; non-Fermi liquid behavior is driven by the
infinitely long-range and retarded interactions that develop between
quasiparticles at the QCP. From the second perspective,
quasiparticles in the Fermi liquid  are composite
bound-states formed in {\sl real space} between  local moments and conduction electrons;
at the critical point,
the bound-states which characterize the Kondo lattice
disintegrate to reveal an underlying lattice of ordered
magnetic moments. The challenge is to reconcile
hese two viewpoints into a single unified picture
of the critical magnetic fluid. 

\subsection{Weak-coupling Approach}

The weak-coupling approach has a long history founded in spin-fluctuation picture
of ferromagnetism.\cite{moriya}  More recently, Hertz and Millis\cite{hertz,millis}
have recast this approach in a modern renormalization-group language. Hertz
made the key observation that at a  QCP, the correlation time diverges
more rapidly than the correlation length, according to the relation
\bea\tau \propto \xi^z, \eea
where $z$ is called the dynamical critical exponent.   Hertz identified
$z$ as an anomalous scaling dimension of time $[\tau] = [x]^z$, raising the effective
dimensionality of a QCP from $3+1$ to $D=3+z$.   When $D>4$, interactions amongst the magnetic
modes vanish in the continuum limit, where they behave as Gaussian degrees of freedom. 
Since spin fluctuation theory predicts $z=2$ at an antiferromagnetic QCP, a priori,
it should provide a perfectly self-consistent description of antiferromagnetic
QCP's in three dimensions. 

In spin-fluctuation theory, the heavy fermion metal
is described
by a featureless  band of electrons, with Hamiltonian
\bea
H = \sum_{\bf k}\epsilon_{\bf k}c\dg_{\bf k \sigma}c_{\bf k \sigma}
+\sum_{{\bf q}}J({\bf q}) {\bf
S}_{\bf q}
\cdot {\bf S}_{- {\bf q}}
\eea
where ${\bf S}_{\bf q} = \sum_{\bf k}c\dg_{{\bf k} }\pmb{$\sigma$} c_{{\bf k}+ {\bf q}}$
is the spin density and $J({\bf q})$ the magnetic interaction strength
 at wavevector ${\bf q}$. 
In an RPA treatment of this model, the dynamical magnetic susceptibility is 
\bea
\chi_({\bf q}, \omega)^{-1} = \chi_o({\bf q}, \omega)^{-1}+ J({\bf q})
\eea
where $\chi_o$ is the bare magnetic susceptibility. 
When interactions reach a  value where, at
some critical wavevector $\bf Q$,   $J({\bf Q})=  -\chi_o({\bf Q})^{-1}
$,  the susceptibility diverges and
magnetic order condenses at  wavevector $\bf Q$. 

\begin{figure}[tb]
% ********   This is for two columns
\epsfxsize=3.0in 
% ***********For one column  ********************
%\epsfxsize=4.0in 
% ***********************************
\centering{\epsfig{figure=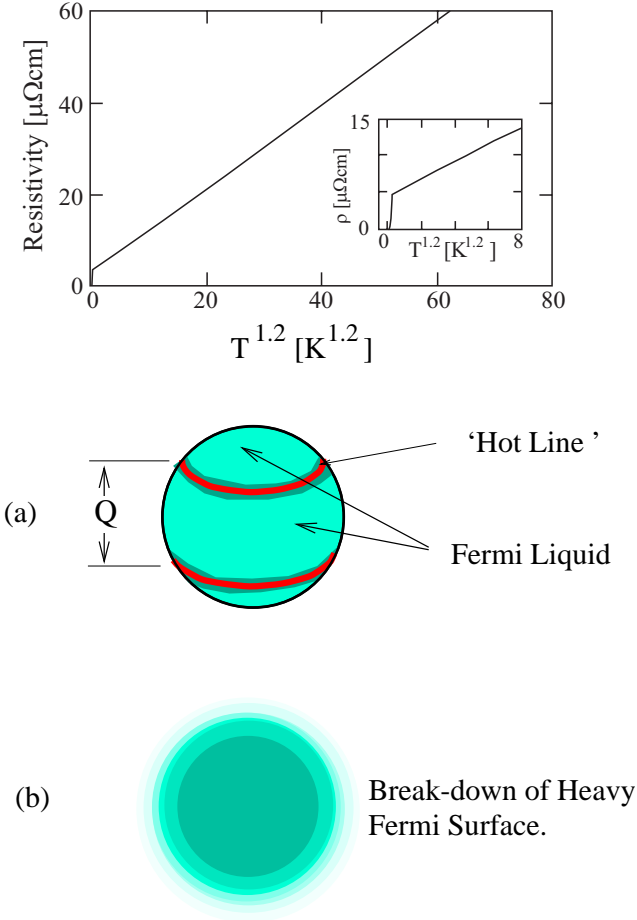,width=\linewidth}}
\vskip 0.1truein
\protect\caption{Anomalous resistivity of $CePd_2Si_2$ at the critical
pressure, after Julian et al.  (a) In the simplest spin-fluctuation
picture, magnetic scattering leads to ''hot lines'' around the Fermi surface,
separated by regions of normal Fermi liquid  which would short-circuit
the conductivity at low temperatures (b) The absence of any $T^2$ regime
in the resistivity  suggests that sharply defined quasiparticles
are eradicated from the entire heavy Fermi
surface.
}
\label{Fig2}
\end{figure}

In the weak-coupling picture, at the QCP,
\bea
\chi_({\bf q}, \omega)^{-1} = 
[\delta + a^2 ({\bf q }- {\bf Q})^2- i \omega /\Gamma ]/\chi_o
\eea
where $\chi_o$ is the uniform magnetic susceptibility, $a$ is of order the
lattice spacing;  $\delta $ is a quantity characterizing the distance from the QCP  and $\Gamma$ 
characterizes the characteristic energy
scale of spin fluctuations.
From this form, we see that the characteristic spin-correlation length and time are given by
\bea
\xi \sim \frac{a}{\delta^{\frac{1}{2}}}, \qquad \tau \sim \frac{1} {\Gamma \delta}
\eea
so that $z=2$. This is one of the \underline{central} predictions
of the weak-coupling approach.
According to the Hertz-Millis theory, in dimensions $d>4-z=2$, it
should be possible to regard the magnetic
fluctuations as a sort of overdamped exchange boson, propagating a long-range
interaction given by
\bea
V(q) = J(q)^2\chi(q , \omega)
\eea
This type of theory  provides a remarkably
good description of 
ferromagnetic QCP's that develop in transition metals.\cite{julian}
Unfortunately, it has become increasingly clear, that it is inadequate
for the QCP in heavy fermion metals. In particular:
\begin{itemize}

\item The theory predicts that the critical  fluctuations
will give rise to an anomalous specific heat
$\Delta C_V/T \propto T^{(d-z)/z}\sim T^{1/2}$ in three dimensions.  
In both $CeNi_2Ge_2$\cite{steglich} and $CeCu_{6-x}Au_x$,\cite{vonlohn}
a logarithmic temperature dependence of $C_V/T$ is observed.

\item The theory does not lead to a break-down of Fermi liquid behavior,
in the strict sense that the quasiparticles are well defined on most
of the Fermi surface, excepting in the vicinity of the ``hot lines''
that are linked by the magnetic $\bf Q$ vector on the Fermi surface (\ref{Fig2}). 
On the hot lines, the quasiparticle scattering rate $\Gamma_{\bf k} \propto \sqrt{T}$,
but away from the hot lines, the quasiparticle scattering rate
has the classic Fermi liquid form $\Gamma_{\bf k } \propto T^2$.\cite{hlubina}

\end{itemize}
This last point
 was first noted by 
Hlubina and Rice,\cite{hlubina} but is still not widely appreciated.  
 The presence of large regions of
normal Fermi surface between the hot lines implies that at temperatures below the
spin-fluctuation scale $\Gamma_o$, the cold regions of the Fermi surface
will short-circuit the conductivity to produce a $T^2$ resistivity.
\noindent 
Such a cross-over to Fermi liquid behavior has never been seen.

These difficulties prompted the Karlsruhe group to carry out a detailed
neutron scattering study of $CeCu_{6-x}Au_x$ at the critical doping $x=0.1$. Early measurements
showed that the spin fluctuation spectrum is  peaked along {\sl rods} in momentum space, rather than
around a single Bragg peak\cite{rosch,stockert}. 
This led Rosch et al. to propose that, the magnetic fluctuations in this material are
quasi-two dimensional.  This hypothesis very elegantly deals with the
two difficulties mentioned above, for once $d=2$, the predicted
specific heat becomes logarithmic; furthermore, the absence of any well
defined scattering wavevector along the length of the magnetic rods,
implies that the hot-rings are smeared  out over a broad region of the
Fermi surface, partially  eliminating the second criticism.  The development
of spin fluctuations of  reduced dimensionality in a fully three-dimensional crystal
is already a very serious departure from our expectations.  Even this is
not enough to save the weak coupling picture, as I now discuss.

\subsection{Strong-coupling Approach}

What could possibly go wrong with the weak-coupling spin fluctuation theory?
One potential 
shortcoming is that it fails to account for the essentially local physics  of the
Kondo effect and moment formation. 
More that two decades ago, Larkin and
Melnikov\cite{melnikov} pointed out that the Kondo effect
can break down at a QCP.  These authors found
that  if a local moment is immersed in a conduction sea at a Ferromagnetic QCP, 
the conventional Kondo effect does not take place; 
instead, the local
moment forms a bound-state with the magnetic lattice in an effect which
does not depend on the sign of the coupling. 

Were the 
 Kondo effect to fail at an
antiferromagnetic QCP, then  the entire strong-coupling
picture of the antiferromagnetic QCP would change.
At present, there is no theory for the Kondo effect at an antiferromagnetic QCP.
However,the  break-down of the Kondo effect at a QCP has been considered in  other
contexts\cite{aronson}, particularly that of quantum spin-glasses.\cite{sachdev,sengupta,parcollet}  A
common thread to these theories is the idea that if the local moment becomes
 critically screened at the QCP, the spin correlations will develop power-law
correlations in time, up to cut-off times of order $\hbar/k_B T$.  This would then lead to
  $\omega/T$ scaling in the spin-susceptibility\bea
\chi(\omega, T) = \frac{1}{T^{\alpha}}f(\frac{\omega}{T})\label{lab1}
\eea
This  kind of behavior is not generally expected in
Hertz-Millis  theory, but apparently,  it {\sl is} observed in heavy fermion systems.
$\omega/T$ scaling was first observed in  neutron scattering experiments
on powder samples of $UCu_4Pd$, which is believed to lie close to a magnetic
QCP.\cite{aronson}  More recently, $\omega/T$ scaling was observed
at the critical wavevector in $CeCu_{5.9}Au_{0.1}$\cite{schroeder}, where the measured magnetic
susceptibility can be fitted by a form
\bea
\chi({\bf q}, \omega)^{-1} = [ J({\bf q}) +a(-i \omega +  T)^{\alpha}], 
\quad(\alpha = 0.8)\label{lab2}
\eea
At the critical  wave-vector $\bf q = \bf Q$, $J({\bf Q})=0 $ vanishes, and (\ref{lab2}) becomes
identical to (\ref{lab1}).  
\begin{figure}[tb]
% ********   This is for two columns
\epsfxsize=3.0in 
% ***********For one column  ********************
%\epsfxsize=4.0in 
% ***********************************8
\rightline{\epsfig{figure=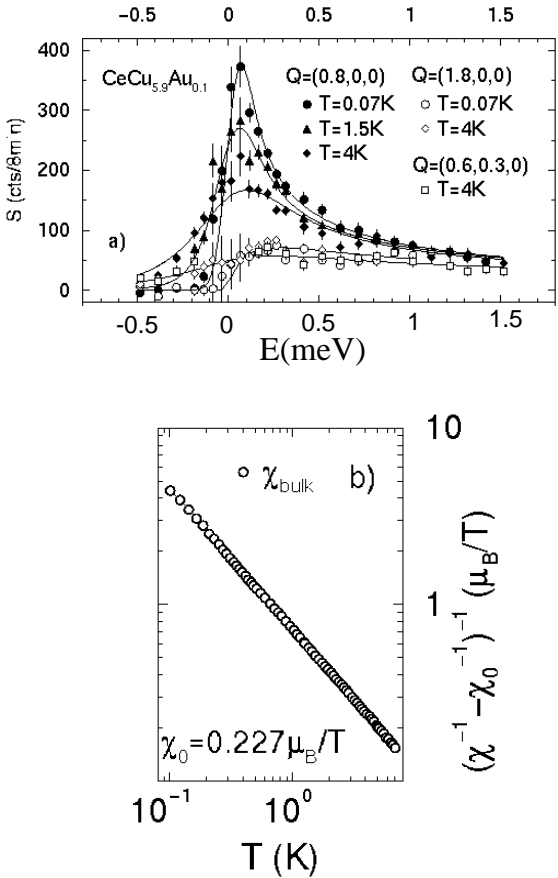,width= 3.0truein}}
\vskip 0.1truein
\protect\caption{$\omega/T$ scaling is manifested
in the inelastic neutron scattering from $CeCu_{6-x}Au_x$, and the very same
scaling exponent is found in the uniform suscepibtility.}
\label{Fig4}
\end{figure}\noindent
Remarkably, the same exponent $\alpha=0.8$ appears  to dominate the 
{\sl entire} Brillouin zone, and re-appears in the uniform
susceptibility, which can be fitted to a form $\chi(T)^{-1} = (J(0) + a T^{0.8})$ (Fig.
\ref{Fig4}) . It follows that the dynamical critical exponent is $2/\alpha = 2.5$- a  serious
departure from the weak-coupling picture.

Sengupta has suggested a simple  way to think about
this type of  behavior, proposing that when the Kondo effect breaks down,
the Weiss field $\vec h$  
acting on the  local moments develops critical correlations in time,
\bea
\langle h_a(\tau) h_b(\tau')\rangle \sim \frac{\delta_{ab}}{(\tau-\tau')^{2-\epsilon}}
\eea
Any value of $\epsilon>0$ represents a strong departure from Fermi liquid behavior.
To see that kind of behavior is   consistent with the magnetic correlations
observed in $CeCu_{6-x}Au_x$, let us  relate the Weiss field to the spin
density by writing $h(q) = J(q) S(q)$. It follows that the Weiss-field correlation
function is 
\bea
\langle h^a({\bf q}, \omega)h^{b}(-{\bf q},- \omega) \rangle 
= J({\bf q})^2 \chi^{ab}({\bf q}, \omega)
\eea 
But if we use the form (\ref{lab2}), the  local correlation function of the
Weiss-field is given by
\bea
\langle h^a( \omega)h^{b}(- \omega) \rangle &=& \delta^{ab}
\int \frac{d^3 q}{(2 \pi)^3}
 \frac{J({\bf q})^2}{J({\bf q}) + a(-i \omega +T)^{\alpha}}\cr
&\propto& (-i \omega +T)^{\alpha}+ {\rm cons}
\eea
where the zero $J({\bf Q})=0$ suppresses the singular contribution
around $\bf q=\bf Q$.   But if ${\rm Im} \langle h^a( \omega)h^{b}(- \omega) \rangle 
\sim \omega^{\alpha}$, it follows that 
$\langle h^a( t)h^{b}(0) \rangle 
\sim 1/t^{1+\alpha}$, showing that 
 $\epsilon= 1 - \alpha =0.2$ in $CeCu_{6-x}Au_x$ .  Notice incidentally  that in spin-fluctuation
theory $\alpha =1$, so $\epsilon=0$ and the Weiss field shows no departure from Fermi liquid
behavior.

Inspired by early work of Sachdev and Ye on quantum spin-glasses\cite{sachdev},
Sengupta\cite{sengupta} has argued that the effective action of a local moment
subjected to  this fluctuating field in a cavity should take the form
\bea
I = \int_0^{\beta} d \tau d\tau'  S^a( \tau) \langle h_a(\tau) h_b(\tau')\rangle S^b( \tau'), 
\eea
At certain stable values of $\epsilon$, the spins
must  develop powerlaw correlations in such a way that the action is
scale-invariant.  Since the scaling dimension of $[h]= \tau^{-1 + \epsilon/2}$,
it follows that $[S]= \tau^{ - \epsilon/2}$, so that 
$\langle S(\tau) S(0) \rangle \sim \tau^{- \epsilon}$, corresponding to
a cavity spin susceptibility $\chi_o(\omega)^{-1}  \sim (-i\omega)^{1 - \epsilon}$.
When we couple the cavity up to its environment, 
\cite{gabi}  we obtain
\bea
\chi({\bf q}, \omega)^{-1} = (\chi_o(\omega )^{-1}+J({\bf q})),\eea
which recovers  the phenomenological form (\ref{lab2}) observed in experiment.
Unfortunately, although these arguements establish 
the stability of a $QCP$ with Weiss fields that are critically
correlated in time, they provide no insight into its microscopic origins, in particular:
(i) the observed value of $\epsilon$, (ii)
the detailed momentum dependence of the magnetic fluctuations,
and  (iii) how the critical magnetic fluctuations couple to, and scatter electrons.

\subsection{More mysteries indicate the need for  new   microscopic theory.}

\newcommand\joinreldex{\mathrel{\mkern-9mu}}
\newcommand\joinrelwx{\mathrel{\mkern-6mu}}

It turns out, as Stockert will discuss at length, that the soft
spin-fluctuations at the QCP actually occupy a ``dog-bone'' shaped
($>\joinrelwx\relbd \joinrelwx \relbd \joinrelwx<$)
region of momentum space, rather than a simple tube as first supposed.\cite{stockert,schroeder}
This is  reminiscent of a frustrated magnet lying close to an incommensurate-
commensurate phase transition, or Lifschitz point.
 This led us\cite{schroeder} to propose that in the vicinity of the magnetic Bragg peak, the
spin-stiffness becomes quartic in momentum in directions parallel to the arms of the ``X''-shaped
scattering maximum, i.e.  $J({\bf q} )\sim A \delta q_{\perp}^2 + B \delta q_{\parallel}^4$.  
In the
Free energy, the soft  directions provide less phase space
for fluctuations and only count as one half-dimension, so the
effective dimensionality for the thermodynamics is $D_{eff}= 2 +\frac{1}{2}$   Curiously, this
exactly  matches the dynamical critical exponent $z=2.5$ deduced from the neutron scattering, so
that a logarithmic specific heat is predicted by the theory, but now the hot-line problem
returns.

In this paper I have tried to argue that the  observation of 
\begin{itemize}

\item critical fluctuations in the Weiss field.

\item Reduced dimensionality of the magnetic fluctuations.

\item Non-trivial exponents in the resistance

\end{itemize}
demand a new approach to the physics of the antiferromagnetic QCP in
heavy fermion systems, possibly one that incorporates the physics of moment
formation and the Kondo effect.  
What sort of features might we expect in such a theory? Let me make a few speculative
remarks to stimulate discussion.
One of the weak
central assumptions of spin-fluctuation theory, is that the amplitude vertex for  the process
\bea
e^-_{{\bf k}\uparrow} \rightleftharpoons e^-_{{\bf k}-{\bf Q}\downarrow}+ \hbox{spin-fluctuation}
\eea
is non-singular at the QCP.  If indeed the Kondo effect fails at
a QCP, it is tempting to
suggest that this assumption must fail. How? Clearly we need to think
about how the composite heavy electron, itself a bound-state
between electrons and local moments, decays. To obtain  a singular
decay process, perhaps one needs some kind of local excitation, giving
rise to a process of the following form
\bea
{\rm f}_{{\bf k}\sigma} \rightleftharpoons {\rm b}_{{\bf k}'\sigma} +
\Phi
\eea
where ${\rm f}_{{\bf k}\sigma} $ represents  the composite heavy electron
formed as part of the Kondo effect, $ {\rm b}_{{\bf k}'\sigma}$
represents the magnetic excitation, written here in the language of Schwinger bosons,
and $\Phi$ is a  massless fermionic excitation (charged, yet spinless)
that has no overlap with the heavy quasiparticle states, which
plays a role in mediating the interaction between heavy fermions
and paramagnons. 
Such a hypothetical excitation is motivated by the observation
of a powerlaw spin correlation that dominates the entire
Brillouin zone; it would also provide a mechanism to destroy the Fermi surface uniformly, rather
than along hot-lines.
This kind of structure may  develop in ``supersymmetric'' approaches
to the Kondo lattice problem, which describe the local moments by
a partner of  bosonic and fermionic degress of freedom.\cite{gan,pepin}

Putting speculations aside, there  is clearly  strong
motivation to extend the sort of measurements made on $CeCu_{6-x}Au_x$
to stoichiometric compounds, such as $CeGe_2Ni_2$, to establish whether
critical Weiss fields are a pervasive feature of the heavy fermion QCP.  On the
theoretical front, there is evidently a desperate need for us
to re-examine the whole issue of how
the local moment starts to reveal itself at quantum criticality.

I should like to thank Gabriel Aeppli,
Alexei Tsvelik, Revaz Ramazashvili,
Gilbert Lonzarich, Antoine Georges, Catherine P\' epin and Achim Rosch for discussions, and
collaborations that have contributed substantially to this paper. 
This work was supported by NSF grant DMR 96-14999.

\end{document}